\documentclass[%
 reprint,
 amsmath,amssymb,
 aps,
]{revtex4-2}
  
\usepackage{graphicx}
\usepackage{dcolumn}
\usepackage{bm}

\usepackage[caption=false]{subfig}

\usepackage{url}
\usepackage{braket}
\usepackage{hyperref}
\hypersetup{
  colorlinks   = true,    
  urlcolor     = blue,    
  linkcolor    = blue,    
  citecolor    = blue,    
}

\usepackage{amsmath} 

\begin{document}

\preprint{APS/123-QED}

\title{Modelling and optimization of pulsed squeezed state generation in a ring-resonator system} 

\author{Marc M. Dignam}
\email{dignam@queensu.ca}
\affiliation{Department of Physics, Engineering Physics \& Astronomy,\\ Queen’s University, Kingston, Ontario K7L 3N6, Canada.}
\author{Marco Liscidini}
\affiliation{Department of Physics, University of Pavia, Pavia, Italy.}

\date{\today} 
\begin{abstract}
We present a semi-analytic formalism for calculating the squeezing and antisqueezing spectrum in a channel waveguide
side-coupled to a lossy ring resonator. Our approach first uses the
semi-analytic evolution of the density matrix inside the ring up to the time
where the squeezing is maximized. Then noting that a
conservative approximate result for the squeezing can be obtained by ignoring the effect
of the pump at later times, we calculate the free evolution of the field
operators in the channel waveguide at all later times. We then
calculate the quadrature squeezing spectrum in the waveguide, assuming that the
measurement starts at the time when the squeezing in the ring is maximized.
Using these results, we determine the optimum values for the pump pulse
duration and amplitude and the ring-channel coupling for the pump and signal. \ We find that squeezing above 10 dB can be easily achieved in the channel for antisqueezing levels of less than 22 dB.
\end{abstract}

\maketitle


\section{Introduction}

The generation of squeezed states, in which the uncertainty in one quadrature of the electromagnetic field is reduced below the vacuum limit, plays a crucial role in many quantum technologies, with applications spanning quantum sensing, secure communication, and computing. For instance, squeezed states can be used to enhance the sensitivity in interferometric setups for gravitational wave detection \cite{Caves1981GravitationalWave}, it enables the encoding of information for quantum communication, providing a pathway toward secure quantum key distribution \cite{Ralph2000QuantumComm}, and has been proposed as a means of implementing continuous-variable quantum gates \cite{Lloyd1999QuantumComp}.

Second- and third-order nonlinear processes, such as spontaneous parametric down-conversion (SPDC) and spontaneous four-wave mixing (SFWM), are particularly effective in generating bright squeezed states, and they can be implemented in photonic integrated devices, with benefits in terms of scalability, efficiency, and control of the generated states. In this respect, integrated resonators in various platforms, ranging from CMOS-compatible materials to thin-film lithium niobate (TFLN), offer significant advantages by enabling strong field confinement 
\cite{Caspani:2017int,Loncar2019TFLN}, thus expanding the potential for quantum photonic technologies.

In any system used to generate a squeezed state of light propagating on-chip, in an integrated waveguide, or in optical fiber, the aim generally is to achieve strong squeezing in one quadrature and minimal antisqueezing in the orthogonal quadrature. If one exploits a nonlinear optical parametric process in an integrated resonator, optimizing such a system requires careful consideration of several key parameters, including (1) the pump pulse duration, (2) the pump pulse energy, (3) the intrinsic and loaded quality factors for the pump in the resonant cavity, and (4) the intrinsic and loaded quality factors for the signal mode in which light is generated.

Modeling SPDC and SFWM in an integrated structure faces several challenges, from field quantization in nontrivial geometries to the description of the system's nonlinear dynamic in the presence of a strong pump field. To this end, over the years, a range of approaches has been developed. In the low-gain regime, perturbative approaches allow for the calculation of the quantum state itself  \cite{Yang2008SPDC, Liscidini2012Asymptotic}, but their accuracy is reduced for bright squeezed states, and they are difficult to implement in the case of significant self- and cross-phase modulation. For higher gain scenarios, alternative methods have been developed, allowing for accurate predictions of squeezing levels. In this case, the focus is usually the expectation values of key observables rather than the full quantum state \cite{Quesada:22, Vernon2015, Vendromin2020}.

In this work, we present a semianalytic method to calculate the quadrature squeezing spectrum in a channel waveguide side-coupled to a ring resonator waveguide. Using this model, we optimize the pump pulse duration and amplitude, and the ring-channel coupling for the pump and signal for given intrinsic ring resonator quality factors. Our approach involves first calculating the evolution of the density matrix inside the ring up to the time at which the squeezing is maximized. We then approximate the squeezing spectrum in the channel waveguide by adopting a conservative estimate that neglects the pump’s effect for later times, allowing the free evolution of the field operators in the ring and channel waveguides. The quadrature squeezing spectrum in the channel waveguide is calculated for a measurement that starts when the ring’s squeezing is maximized. As an example, we find that in a TFLN ring with an intrinsic quality factor ($Q_{sI}$) of 2,000,000 at a signal wavelength of 1550 nm and an intrinsic $Q_{pI}$ of 800,000 at the pump wavelength, a squeezing level exceeding 10 dB can be achieved in the channel, with only 22 dB of antisqueezing for a pump duration of approximately 2.3 ps. 

The paper is organized as follows. In Sec. \ref{Theoretical model}, we present our theoretical model. This begins in Sec. \ref{Overview} with an overview of our approach, which is followed by a discussion of the Hamiltonian (Sec. \ref{Hamiltonian}), the solution to the Lindblad master equation (Sec. \ref{Lindblad}), and the squeezing spectrum (Sec. \ref{Squeezing}). In Sec. \ref{Results}, we present the result of our calculations and we finish with our conclusions in Sec. \ref{conclusion}.

\section{Theoretical model} \label{Theoretical model}
\subsection{Overview} \label{Overview}

\begin{figure}[hbt]
\includegraphics[width=\columnwidth]{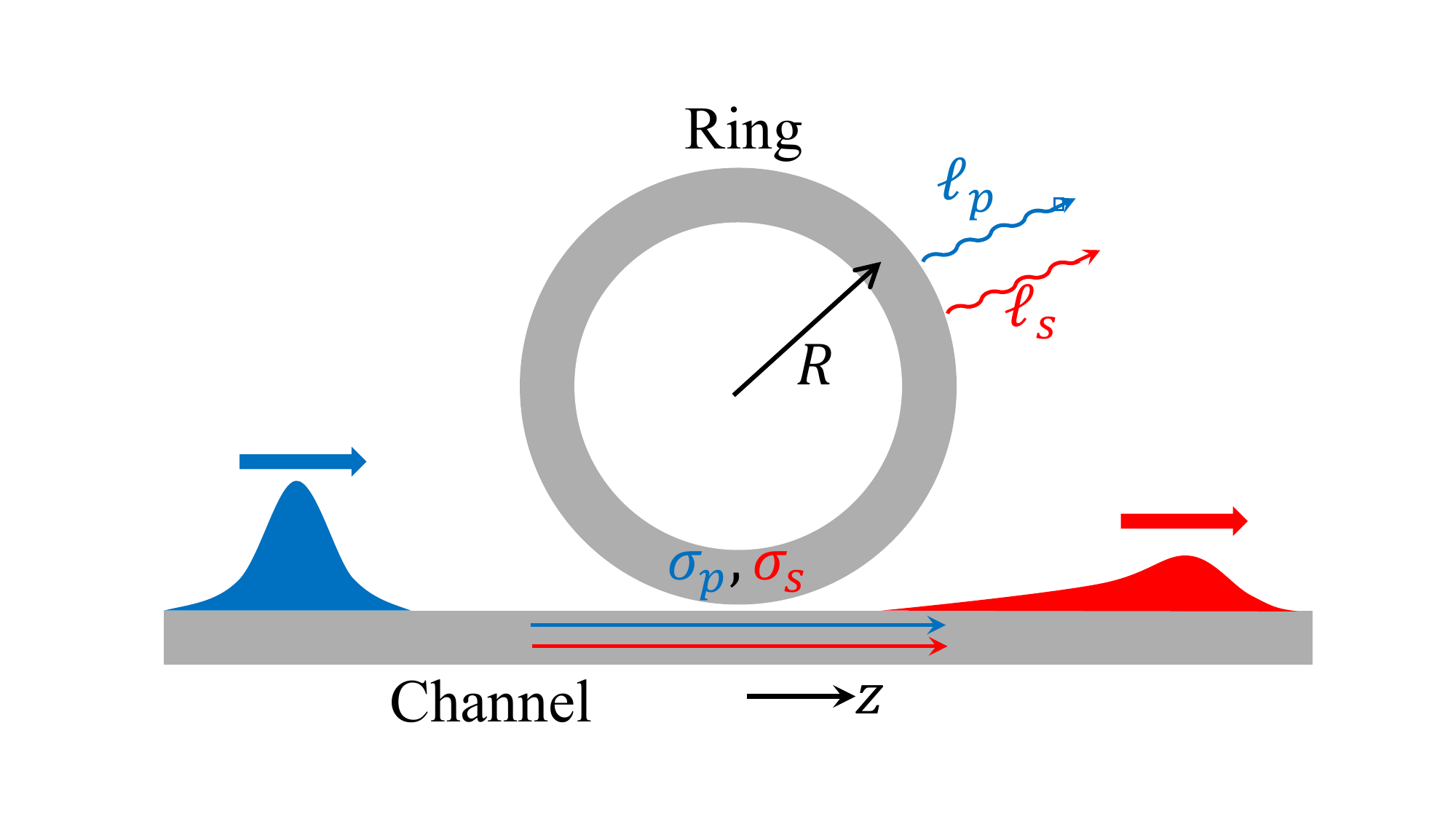}
\caption{Schematic diagram of the microring resonator side-coupled to a channel waveguide. The ring has a radius $R$, with intrinsic scattering losses at the signal and pump frequencies, that lead to round-trip field loss factors $\ell_s$ and $\ell_p$, respectively. The coupling of the ring to the channel is such that the through-coupling coefficient is $\sigma_p$ and $\sigma_s$ at the signal and pump frequencies, respectively. 
 Also shown schematically is the Gaussian pump pulse entering heading towards the ring and the pulse of signal light that exits the ring. The coupling is taken to be a point coupling at $z=0$ in the channel waveguide}\label{fig:ringresonator}
\end{figure}

The system we are considering is shown schematically in Fig. 1.
It consists of a microring resonator of radius $R$, side coupled to a straight channel waveguide. Both waveguides are made from a nonlinear material (such as Lithium Niobate or AlGaAs) with a significant $\chi^{(2)}$ response. The pump field is a pulsed, coherent state, with a center frequency $\omega_P$ that is initially propagating in the channel waveguide. The pump frequency is resonant with one of the modes of the ring resonator and its bandwidth is sufficiently narrow that it only couples to that mode.  Inside the ring, the pump field amplitude is enhanced, and we assume that it is only inside the ring that the pump is strong enough to produce a significant nonlinear response.  In the ring, the process of spontaneous parametric down-conversion results in the generation of squeezed light at the signal frequency, $\omega_s=\omega_p/2$. This squeezed light is then coupled out of the ring back into the channel waveguide, wherein the squeezing spectrum is evaluated. 

In this section, we present the multi-step theoretical approach that we employ to calculate
the squeezing spectrum of the quadrature of the light in the channel waveguide. The key steps in this approach
are as follows. First, we determine the approximate but very accurate
expression for the optical pump field in the ring for an incident
temporally Gaussian pulse. Second, we use the classical pump field in the ring
along with the material nonlinear response to calculate the density operator
of the squeezed signal light in the ring as a function of time. We use this result to
determine the time $t_{m}$, at which the quadrature is maximally squeezed in
the ring.  Because the squeezing will always be better the longer the pump
lasts, we make the conservative approximation that we can ignore
the pump field for times greater than $t_{m}$. We then derive analytic
expressions for the two-time correlation functions for the light in the ring
for times greater than $t_{m}$; using these expressions, we are able to derive
semianalytic expressions for the squeezing spectrum in the waveguide for
measurements starting at $t=t_{m}$.

The discussion of the theoretical approach is organized as follows.  In Sec. B we present the Hamiltonian describing the light in the ring along with the closed-form solution for the corresponding density operator.  In Sec. C, we derive the approximate expressions for the two-time correlation functions for the light in the ring for times after the maximal squeezing has occurred. Finally, in Sec. D we derive the results for the squeezing spectrum in the channel waveguide and relate this to the two-time correlation functions in the ring.

\subsection{The signal Hamiltonian and pump field} \label{Hamiltonian}

In this section, we present the Hamiltonian for
the quantum state of light of the signal in a single mode of the ring resonator as created
through the nonlinear process of degenerate parametric down conversion. Considering a material with a nonlinear $\chi^{\left(  2\right)  }$
response, a single photon of frequency $\omega_{p}$ is converted into two
signal photons of frequency $\omega_s=\omega_{p}/2$. The number of photons in
the pump is assumed to be much larger than the number of nonlinearly generated
signal photons, so we approximate the pump field in the pump mode in the ring as an undepleted coherent state, with a time-dependent coherent state complex amplitude, $\alpha_p(t)$.
This leads to the Hamiltonian \cite{Quesada:22}
\begin{equation}
\hat{H}=\hbar\omega_s\widehat{a}^{\dagger}\widehat{a}+\eta^{\ast}%
\alpha_{p}^{\ast}\left(  t\right)  \widehat{a}\widehat{a}+\eta \alpha_{p}%
\left(  t\right)  \widehat{a}^{\dag}\widehat{a}^{\dag},
\label{EqHam}
\end{equation}
where $\widehat{a}$ ($\widehat{a}^{\dag}$) is the annihilation
(creation) operator for signal photons in the resonant ring mode, and  
\begin{equation}
   \eta \equiv -\hbar \omega_s\chi^{(2)}_{eff}\sqrt{\frac{\hbar \omega_p}{16\pi\epsilon_o R A_{eff}}} 
\end{equation}
is the nonlinear coupling coefficient between the pump and signal
modes, where $\chi^{(2)}_{eff}$ is an effective
nonlinear susceptibility that depends on the intrinsic nonlinear susceptibility of the
ring material and spatial mode profiles in the ring, and $A_{eff}$ is the effective transverse area of the ring waveguide modes \cite{Quesada:22}. 

The frequencies of the modes in the ring resonator are given by
\begin{equation}
\omega_m=\frac{2\pi m}{T_R(\omega_m)},    
\end{equation}
where
\begin{equation}
    T_R(\omega)\equiv\frac{n_{eff}(\omega)2\pi R}{c}
\end{equation}
is the round trip time for the $m^{th}$ mode, $n_{eff}(\omega)$ is the effective index of refraction in the ring waveguide at frequency $\omega$, and $R$ is the ring radius.
Due to
intrinsic losses and coupling to the straight waveguide, the ring dwell time of the pump and signal photons in the ring is finite.
The intrinsic amplitude loss factor per round trip in the ring for the signal (pump) is denoted
by $\ell_{s}$ ($\ell_{p}$) and the signal (pump) self-coupling of the ring to the straight waveguide
is denoted by $\sigma_{s}$ ($\sigma_{p}$). We let the complex frequency of the signal/pump modes in
the ring be%
\begin{equation}
\Omega_{s(p)}=\omega_{s(p)}-i\Gamma_{s,(p)}/2,
\end{equation}
where the loaded power loss coefficient can be expressed approximately as%
\begin{align}
\Gamma_{s(p)}  &  =-\frac{2}{T_{Rs(p)}}\ln{\left(\sigma_{s(p)}\ell_{s(p)}\right)}\nonumber\\
&  \simeq\frac{2\left(  1-\sigma_{s(p)}\ell_{s(p)}\right)  }{T_{Rs(p)}},\label{EqnGamma}
\end{align}
where $T_{Rs(p)}\equiv T_R(\omega_{s(p)})$ and
the second line is a good approximation as long as we are in the high-finesse regime, which will be the case for for the signal, but not necessarily for the pump. The intrinsic loss coefficients are calculted from Eq. (\ref{EqnGamma}), with $\sigma_{s(p)}=1$.
The intrinsic and loaded quality factors for the signal and pump are $Q_{sI}=\omega_s/\Gamma_{sI}$ and $Q_{pI}=\omega_p/\Gamma_{pI}$, respectively, while the loaded quality factors are $Q_{sL}=\omega_s/\Gamma_{sL}$ and $Q_{pL}=\omega_p/\Gamma_{pL}$.

In all
that follows, we assume that the pump and signal are on-resonance with ring
modes, and the pump is not chirped, such that 
\begin{equation}
\eta \alpha_{p}\left(
t\right)  =\left\vert \eta \alpha_{p}\left(  t\right)  \right\vert e^{i\theta
}e^{-i2\omega_s t},
\end{equation} 
where $\theta$ is independent of time.

We take the incident pump field in the
channel waveguide to be a temporally Gaussian pump, with an intensity full-width at half maximum (FWHM) $\tau_{p}$ that peaks at the coupling point to the ring at time $t=0$. Thus, expanding  in the channel waveguide longitudinal modes, the pump field in the channel for $z<0$ is given by
\begin{align}
    \mathbf{E}_c(\mathbf{r};t)=i\sqrt{\frac{\hbar
    \omega_p}{2\epsilon_o}}\mathbf{F}(x,y)\int dk \widetilde{\alpha}_{in}(k)\frac{e^{i(kz-\omega_c(k)t)}
    }{\sqrt{2\pi}}  + c.c.,
\end{align}
where $\widetilde{\alpha}_{in}(k)$ is the coherent state amplitude for the mode with wavevector $k$, $\omega_c(k)$ is the frequency dispersion in the channel waveguide, and $\mathbf{F}(x,y)$ is the transverse mode, which is normalized such that
\begin{equation}
\int dxdy\epsilon_r (x,y)  |\mathbf{F}(x,y)|^2=1,  
\end{equation}
where $\epsilon_r (x,y)$ is the transverse profile of the dielectric constant.
Now, we take the coherent pump amplitude in the channel to be
\begin{equation}
\widetilde{\alpha}_{in}\left( k\right)  =\widetilde{\alpha}_{o}
\exp\left[  -\frac{(k-k_p)^2}{\kappa^2
}\right],    
\end{equation}
where $\kappa$ is the pulse width parameter. Thus, using the approximate dispersion function, $\omega_c(k)=\omega_p + v_{gp}(k-k_p)$, where $v_{gp}$ is the group velocity at the pump frequency, the pump field in the channel waveguide just before the coupling point ($z=0^-$) becomes
\begin{align}
    \mathbf{E}_c(x,y,0^-;t)=& i\sqrt{\frac{\hbar
    \omega_p}{2\epsilon_o}}\frac{2\sqrt{\ln (2)}}{\tau_p v_{gp}}\mathbf{F}(x,y)\widetilde{\alpha}_{o}\nonumber\\
    &\times \exp [-2 \ln(2)t^2/\tau_p^2]e^{-i\omega_pt}
      + c.c.,
\end{align}
where $\tau_p$ is the temporal FWHM of the pump pulse in the channel waveguide, which is related to $\kappa$ by
\begin{equation}
    \kappa \equiv \frac{\sqrt{8\ln{2}}}{\tau_p v_{gp}}.
\end{equation}
Using the above expressions, it is easily shown that the number of photons in the channel waveguide is given by
\begin{equation}
    N_c=\sqrt{\frac{4\ln(2)}{\pi}}\frac{T_{Rp}|\widetilde{\alpha_o}|^2}{\tau_pR}.
\end{equation}

Now, the pump field in the ring just past the coupling point can be written as
\begin{align}
    \mathbf{E}_R(x,y,0^+;t)=& i\sqrt{\frac{\hbar
    \omega_p}{2\epsilon_o}}\mathbf{F}(x,y)\alpha_{p}(t)\frac{e^{-i\omega_pt}}{\sqrt{2\pi R}} + c.c.,
\end{align}
where $\alpha_{p}(t)$ is the coherent state amplitude for the pump in the ring. The Fourier transform of the coherent amplitude in the ring is related to the Fourier transform of channel amplitude by \cite{Vendromin2020}
\begin{equation}
    \widetilde{\alpha}_p(\omega)=i\frac{\sqrt{1-\sigma_p^2}\ell_pe^{i\omega T_R}}{1-\sigma_p \ell_p e^{i\omega T_R}}\frac{2\pi\sqrt{R}}{v_{gp}}\widetilde{\alpha}_{in}(k_p +(\omega - \omega_p)/v_{gp}).
    \label{EqAlphap}
\end{equation}

It can be shown that as long as
$\left(  1-\sigma_{p}\ell_{p}\right)  \ll1$, then to a very good approximation, the Fourier transform of Eq. (\ref{EqAlphap}) is given by \cite{Vendromin2020}
\begin{align}
\alpha_{p}\left(  \widetilde{t}\right)    = & i\sqrt{\frac{N_c\widetilde{\tau}_p(1-\sigma_p^2) \ell_{p}^2}{8}}\left[\frac{\pi}{\ln(2)}\right]^{\frac{1}{4}}\operatorname{erfc}
\left[  y\left(  \widetilde{t}\right)  \right]\nonumber\\ 
&\times \exp{\left[  y^2\left(\widetilde{t}\right)-2\ln{(2)}\widetilde{t}^2/\widetilde{\tau}_p^2\right]} ,        
\end{align}
where $\widetilde{t}\equiv t/T_{Rp}$, $\widetilde{\tau_{p}}\equiv \tau_{p}/T_{Rp}$,%
\[
y\left(  \widetilde{t}\right)  \equiv\frac{\left(  1-\sigma_{p}\ell_{p}\right)
\widetilde{\tau}_{p}}{\sqrt{8\ln\left(  2\right)  }}-\frac{\sqrt{8\ln\left(
2\right)  }\widetilde{t}}{2\widetilde{\tau}_{p}},
\]
and $\operatorname{erfc}\left(  x\right)  $ is the complementary
error function. The function $\alpha_{p}\left(  \widetilde{t}\right)$, is pump function that appears in Eq. (\ref{EqHam}) for the nonlinear Hamiltonian. We find that this expression differs from the exact expression by less than 10\%, even for a pump finesse as low as 20, 
and so we will use it in all that follows.  


\subsection{Solution to the Lindblad master equation} \label{Lindblad}

The Lindblad master equation that governs the evolution of the density operator for the quantum state of the signal in the ring is given by
\begin{equation}
        \frac{d\widehat{\rho}}{dt} = -\frac{i}{\hbar}\left[\widehat{H},\widehat{\rho}\right] + \Gamma_{sL} \left[\widehat{a}\widehat{\rho}\widehat{a}^{\dagger} - \frac{1}{2}\left\{{\widehat{a}^{\dagger}\widehat{a}},{\widehat{\rho}}\right\}\right],
    \label{master}
    \end{equation}
where $\{\widehat{A},\widehat{B}\}$ is the anticommutator.    
It has been shown by Seifoory \textit{et al.} \cite{SeifooryJOSAB} that the exact solution to Eq. (\ref{master}) is the squeezed thermal state,
\begin{equation}
\hat{\rho}\left(  t\right)  =\hat{S}\left(  z\left(  t\right)  \right)
\hat{\rho}_{T}\left(  n_{th}\left(  t\right)  \right)  \hat{S}^{\dagger
}\left(  z\left(  t\right)  \right),
\end{equation}
where%
\begin{equation}
\hat{S}\left(  z\left(  t\right)  \right)  \equiv\exp\left\{  \frac{1}%
{2}\left[  z^{\ast}\left(  t\right)  \widehat{a}\widehat{a}-z\left(  t\right)
\widehat{a}^{\dag}\widehat{a}^{\dag}\right]  \right\}
\end{equation}
is the squeezing operator with%
\begin{equation}
z\left(  t\right)  \equiv r\left(  t\right)  e^{i\phi\left(  t\right)  },
\end{equation}
and%
\begin{equation}
\hat{\rho}_{T}\left(  n_{th}\right)  \equiv\frac{1}{n_{th}+1}\left(
\frac{n_{th}}{n_{th}+1}\right)  ^{\widehat{a}^{\dag}\widehat{a}}%
\end{equation}
is a thermal state with an average of $n_{th}$ photons. This is the solution as long as the time-dependent squeezing parameter and the thermal state number obey the following the differential equations:

\begin{align}
\frac{dr}{dt}=&-\frac{2}{\hbar}\left\vert \eta \alpha_{p}\left(  t\right)
\right\vert \sin\left[  \theta-2\omega t-\phi\left(  t\right)  \right] \nonumber\\
&-\frac{\Gamma_{sL}\cosh\left[  r\left(  t\right)  \right]  \sinh\left[
r\left(  t\right)  \right]  }{2n_{th}\left(  t\right)  +1},
\end{align}%
\begin{equation}
\frac{d\phi}{dt}=-2\omega+\frac{2\left\vert \eta \alpha_{p}\left(  t\right)
\right\vert \cos\left[  \theta-2\omega t-\phi\left(  t\right)  \right]
}{\hbar\tanh\left[  r\left(  t\right)  \right]  },    
\end{equation}

\begin{equation}
\frac{dn_{th}}{dt}=\Gamma_{sL}\left\{  \sinh^{2}\left[  r\left(  t\right)
\right]  -n_{th}\left(  t\right)  \right\}  . \label{eq:thermal differential}%
\end{equation}

To simplify the equations, in the rest of this section, we will take time to be in units of the round-trip time for the pump in the ring; thus, we define $\widetilde{t}\equiv t/T_{Rp}$, $\widetilde{\tau_{p}}\equiv \tau_{p}/T_{Rp}$, $\widetilde{\omega}_{s}\equiv \Gamma_{sL}T_{Rp}$, and $\widetilde{\Gamma}_{sL}\equiv \Gamma_{sL}T_{Rp}$. We take the initial state at time $t=t_{o}$ (where $t_o \ll -\tau_{p}$, so that it is well before the pulse arrives) to be the vacuum state. Then, taking $\phi\left(  t_{o}\right)  =-2\omega_s t_{o}+\theta+\pi/2$, gives 
\begin{equation}
\frac{dr}{d\widetilde{t}}=\widetilde{\Gamma}_{sL}\left[ \frac{g\left(\widetilde{t}\right)}{2} -\frac{\cosh\left(r\right)
\sinh\left(r\right)  }{2n_{th}  +1}\right] ,
\label{Eqrdiff}%
\end{equation}%
and
\begin{equation}
\frac{dn_{th}}{d\widetilde{t}}=\widetilde{\Gamma}_{sL}\left[ \sinh^{2}\left(r\right)
-n_{th}\right],\label{Eqnthdiff}%
\end{equation}
where, 
\begin{align}
g(\widetilde{t}) & \equiv \frac{4\left\vert \eta \alpha_{p}\left(  \widetilde{t}\right)
\right\vert}{\hbar\Gamma_{sL}} \nonumber \\
= & \frac
{g_0\sqrt{\pi\widetilde{\tau_{p}}\left(  1-\sigma_{p}^{2}\right)  \ell_{p}^{2}} e^{\left[  z\left(  \widetilde{t}\right)  \right]  ^{2}}  }{\widetilde{\Gamma}_{sL}\sqrt{8\ln\left(
2\right)  }}\nonumber\\
&\times \operatorname{erfc}
\left[  z\left(  \widetilde{t}\right)  \right]\exp\left[  -2\ln\left(  2\right)  \frac{\widetilde{t}^2}{\widetilde{\tau_{p}}
^{2}}\right],     
\end{align}
where
\begin{equation}
g_0\equiv \frac{4\left\vert \eta \right\vert T_{Rp}\sqrt{N_c}} {\hbar}\left[\frac{\ln(2)}{\pi}\right]^{\frac{1}{4}}.     
\end{equation}
is the dimensionless pump strength.
Also, for all times $\widetilde{t}>\widetilde{t}_{o}$, we have simply
\begin{equation}
\phi\left(  \widetilde{t}\right)     =-2\widetilde{\omega}_s \widetilde{t}+\theta+\pi/2.
\end{equation}

Using Eqs. (\ref{Eqrdiff}) and (\ref{Eqnthdiff}) with the analytic expression
for the pump field, we can calculate the quantum state of the light in the
signal mode in the ring as a function of time. Moreover, it is known that for a squeezed thermal state, the two-operator expectation values are given by \cite{PhysRevA.40.2494}
\begin{equation}
\left\langle \widehat{a}^{\dag}\left(  t\right)  \widehat{a}\left(  t\right)\right\rangle 
 =\left(  n_{th}(t)+\frac{1}{2}\right)  \cosh\left[  2r(t)\right]  -\frac{1}{2}, 
\label{Eqadaga}
\end{equation}
\begin{equation}
\left\langle \widehat{a}\left(  t\right)  \widehat{a}\left(  t\right)\right\rangle 
 =-\left(  n_{th}(t)+\frac{1}{2}\right)  e^{i\phi(t)}\sinh\left[  2r(t)\right].
\label{Eqaa}
\end{equation}

We define the quadrature operators in the ring as
\begin{align}
\widehat{X} & \equiv \widehat{a}^{\dagger}e^{i\beta(t)} + \widehat{a}e^{-i\beta(t)},\\
\widehat{Y} & \equiv -i\left(\widehat{a}^{\dagger}e^{i\beta(t)} - \widehat{a}e^{-i\beta(t)}\right),
\end{align}
where the quadrature phase, $\beta(t)=\phi (t)/2=-\omega_s t + \theta/2 + \pi/4$, is chosen to remove the fast oscillations in time, as would be achieved by a local oscillator when performing homodyne detection.

Using the above equations, we obtain 
\begin{equation}
\Delta X^{2}(t)=\left(  2n_{th}(t)+1\right)  e^{-2r(t)},%
\end{equation}
\begin{equation}
\Delta Y^{2}(t)=\left(  2n_{th}(t)+1\right)  e^{+2r(t)}.%
\end{equation}

Alternatively, one can show that the quadrature variances satisfy the following
differential equations:%
\begin{align}
\frac{d\left[  \Delta X^{2}\right]  }{dt}  & =\Gamma_{sL}\left\{1-\left[
1+g\left( t\right)  \right]  \Delta X^{2}\right\}  ,\label{EaDxevol}\\
\frac{d\left[  \Delta Y^{2}\right]  }{dt}  &  =\Gamma_{sL}\left\{1-\left[
1-g\left( t\right)  \right]  \Delta Y^{2}\right\}  .
\end{align}
Thus the squeezing can be found in closed form by simply integrating a first-order differential equation. From Eq. (\ref{EaDxevol}), it is clear that if the pump function is
set to zero for all times greater than, say, $t_m$, then we are guaranteed that $\Delta X^{2}$ will evolve such that
it is greater than it would have been if the pump were still present. \ \emph{Therefore, making the
approximation that the pump goes to zero after the variance reaches its
minimum will always result in a underestimation of the squeezing}. This is the approach that we will take to determine the two-time correlation functions in Sec. D.  

\subsection{The Squeezing Spectrum} \label{Squeezing}

In the previous section, we calculated the time evolution of the density operator for the light in a single mode in the ring.  We now wish to calculate the squeezing spectrum of the light in the channel waveguide.  As has been discussed previously \cite{Vernon2015}, this calculation is complicated by the fact that the modes in the channel waveguide form a continuum.  Therefore, we cannot simply calculate the squeezing in a single channel mode. Instead, we will use the results from the last section along with the pumpless evolution of the ring operators to calculate the
time-limited squeezing spectrum in the channel waveguide. 

We let $b^{\dagger}(k,t)$ ($b(k,t)$) be the Heisenberg creation (destruction) operator for the light in the channel waveguide mode with wavevector, $k$, where
\begin{equation}
[b(k,t),b^{\dag}(k',t)]=\delta(k-k').
\label{Eqbcomm}
\end{equation}
Following Seifoory \textit{et al.} \cite{Seifoory2021}, we define the channel waveguide field creation (annihilation) operator for light at position $z$ in the waveguide to be
\begin{equation}
\psi^{\dagger}(x,t) \equiv \frac{1}{\sqrt{2\pi}}\int_{k_{s}-\delta k/2}^{k_{s}+\delta k/2} dk b^{\dagger}(k,t) e^{i(k-k_s)z},  
\label{Eqpsidef}
\end{equation}
where $k_s$ is the wavevector of the channel mode that is resonant on the chosen ring-resonator signal mode (\textit{i.e.}, $\omega_c(k_s)=\omega_s$ ), and 
\begin{equation}
\delta k=\frac{2\pi}{T_R(\omega_s)}    
\end{equation}
is wavevector spacing between the channel mode that is resonant with the signal mode in the ring and the wavevector of the mode that is resonant with the ring mode with $m=m_s +1$, \textit{i.e.},
\begin{equation}
\omega_c(k_s+\delta k)=\frac{2\pi (m_s+1)}{T_R(\omega_{m_s+1})}.    
\end{equation}

Following Vernon and Sipe \cite{Vernon2015}, we take the coupling of the ring to the channel waveguide to be a point coupling at $z=0$.  Then, one can show that the field operator in the channel waveguide just to the left of the coupling point is given by
\begin{equation}
\psi\left(  0^{+},t\right)  =\psi\left(  0^{-},t\right)  e^{-i\omega_s
t}-i\frac{\gamma}{v_{gs}}a\left(  t\right)  ,
\end{equation}
where $\psi\left(  0^{-},t\right)  $ is the field
operator in the channel waveguide just to the left of the ring contact,
$v_{gs}$ is the group velocity for the signal in the
channel waveguide and $\gamma$ is a parameter describing the total coupling between the ring and the waveguide modes and has units of $m^{1/2}s^{-1}$. The coupling parameter is related to the intrinsic and loaded Q's of the ring by
\begin{equation}
\frac{\left\vert \gamma\right\vert ^{2}}{v_{gs}\Gamma_{sL}} 
=\left[1-\frac{Q_{sL}}{Q_{sI}}\right].  
\label{Eqgamsq}
\end{equation}

Now, we are considering pulsed pumping of the system and so we wish to determine the measurement of the squeezing spectrum in the channel waveguide starting at the measurement time $t_m$.  Although the definition of the squeezing spectrum in the CW case is clear  \footnote{See, \textit{e.g.}, supplemental material in Ref. \cite{Zhang2021}}, there are a variety of possible ways to define this quantity for a pulsed state.  Some of these depend on the details of the detector, some are time-limited measurements, while others are so-called "instantaneous" power spectra \cite{Cresser1983}.  We will use the instantaneous spectrum as proposed by Page \cite{PagePowerSpect} and Lampard \cite{LampardPowerSpect}, which in our case takes the form
\begin{equation}
S(\Omega)\equiv 2v_{gs} \int_0^{\infty}\langle X_{\beta} (t_m)X_{\beta} (t_m + \tau ) \rangle \cos (\Omega \tau )d\tau ,  
\label{EqSpectrum1}
\end{equation}
where
\begin{equation}
X_\beta (t) \equiv  \widetilde{\psi}
^{\dag}\left(  0^{+},t\right)e^{-i\beta} + \widetilde{\psi}
\left(  0^{+},t\right)e^{i\beta},   
\end{equation}
where
\begin{equation}
\widetilde{\psi}\left(  0^{+},t\right)  \equiv\psi\left(
0^{+},t\right)  e^{i\omega_s (t-t_m)}
\end{equation}
and $\beta$ is the local oscillator phase. Now
\begin{align}
\langle X_{\beta} (t_m) & X_{\beta} (t_m + \tau) \rangle  =  
 \left\langle \widetilde{\psi}
\left(  0^{+},t_{m}\right)  \widetilde{\psi}^{\dag} \left(  0^{+},t_{m}+\tau
\right)  \right\rangle \nonumber\\
& + \left\langle \widetilde{\psi}^{\dag}
\left(  0^{+},t_{m}\right)  \widetilde{\psi} \left(  0^{+},t_{m}+\tau
\right)  \right\rangle \nonumber\\
& +\left\langle \widetilde{\psi}
\left(  0^{+},t_{m}\right)  \widetilde{\psi} \left(  0^{+},t_{m}+\tau
\right)  \right\rangle e^{i2\beta} \nonumber\\
& + \left\langle \widetilde{\psi}^{\dag}
\left(  0^{+},t_{m}\right)  \widetilde{\psi}^{\dag} \left(  0^{+},t_{m}+\tau
\right)  \right\rangle e^{-i2\beta}.
\end{align}

Using Eqs. (\ref{Eqbcomm}) and (\ref{Eqpsidef}), along with the fact that for $z>0$ the Hamiltonian for the field operator in the channel waveguide is simply the free evolution Hamiltonian, we obtain
\begin{align}
[ \widetilde{\psi}(0^+,t),\widetilde{\psi}^{\dag}(0^+,t+\tau)] & =\frac{1}{2\pi}\int_{k_{s}-\delta k/2}^{k_{s}+\delta k/2} dk e^{i\omega_c(k)\tau}\nonumber\\
& \approx \frac{1}{v_{gs}}\delta (\tau),
\end{align}
where we have used $\omega_c(k)\approx \omega_s + v_{gs}(k-k_s)$. The second line is an approximation because the integration is over a finite range of $k$, but is a good approximation for use in Eq. (\ref{EqSpectrum1}) as long as $\Omega T_{Rs} \ll \pi$. Thus, we have
\begin{align}
\langle \widetilde{\psi}(0^+,t_m),\widetilde{\psi}^{\dag}(0^+,t_m+\tau)\rangle & = \frac{1}{v_{gs}}\delta (\tau)\nonumber\\
& + \langle \widetilde{\psi}^{\dag}(0^+,t_m +\tau)\widetilde{\psi}(0^+,t_m)\rangle.    
\end{align}

Since there are no signal photons in the waveguide in the region $z<0$, we obtain
\begin{align}
\langle \psi^{\dag}\left(  0^{+},t_{m}\right) & \psi\left(
0^{+},t_{m}+\tau\right)  \rangle \nonumber\\
&=\frac{\left\vert \gamma\right\vert ^{2}}{v_{g}^{2}}\langle a^{\dag
}\left(  t_{m}\right)  a\left(  t_{m}+\tau\right)  \rangle      
\end{align}
and
\begin{align}
\langle \psi\left(  0^{+},t_{m}\right) & \psi\left(
0^{+},t_{m}+\tau\right)  \rangle \nonumber\\
&=\frac{\left\vert \gamma\right\vert ^{2}}{v_{g}^{2}}\langle a
\left(  t_{m}\right)  a\left(  t_{m}+\tau\right)  \rangle  .    
\end{align}

To proceed, we need to evaluate the two-time correlation functions for the operators in the ring for time $t>t_m$. This
is not possible if one only knows the density operator in the ring. \ One approach would be to find the solution to the adjoint master equation 
\begin{equation}
        \frac{d\widehat{A}}{dt} = -\frac{i}{\hbar}\left[\widehat{A},\widehat{H}\right] + \Gamma_{sL} \left\{\widehat{a}^{\dagger}\widehat{A}\widehat{a} - \frac{1}{2}\left[{\widehat{a}^{\dagger}\widehat{a}},{\widehat{A}}\right]\right\},
    \label{eqnajoint}
    \end{equation}
for the desired operator combinations instead. Unfortunately, we do not have analytic solutions to Eq. (\ref{eqnajoint})
when the pump is present. However, as has previously been shown \cite{MohsenK}, if
the pump is absent, then for any normally-ordered combination of the ring operators $\widehat{a}$ and $\widehat{a}^\dagger$, the solution to the adjoint master equation is simply
\begin{equation}\label{afree}
\widehat{a}(t)=  \widehat{a}(t_m)e^{-i\Omega_s(t-t_m)}.  
\end{equation}
So the evolution of the two two-time correlation functions of interest are given simply by \begin{equation}
\langle \widehat{a}^{\dagger}(t_m+t_1)\widehat{a}(t_m+t_2)\rangle=  \langle \widehat{a}^{\dagger}(t_m)\widehat{a}(t_m)\rangle e^{i\Omega_s^\ast t_1}e^{-i\Omega_s t_2}  
\end{equation}
and
\begin{equation}
\langle \widehat{a}(t_m+t_1)\widehat{a}(t_m+t_2)\rangle=  \langle \widehat{a}(t_m)\widehat{a}(t_m)\rangle e^{-i\Omega_s t_1}e^{-i\Omega_s t_2}.  
\end{equation}
Therefore, to proceed, we assume that after some time
$t_{m}$, we can neglect the effect of the pump on the quantum state of the
signal. Then, we use Eqs. (\ref{Eqadaga}) and (\ref{Eqaa}) as our expressions for the two-time correlation functions for the signal light in the waveguide, and obtain for the squeezing spectrum the simplified expression
\begin{align}
S(\Omega) & = 4v_{gs} \left[ \langle \widehat{a}^{\dagger}(t_m)\widehat{a}(t_m)\rangle +\Re \left\{ \langle \widehat{a}(t_m)\widehat{a}(t_m)\rangle e^{-i2\beta}\right\} \right]\nonumber\\
&\times \int_0^{\infty}e^{-\Gamma_{sL} \tau/2} \cos (\Omega \tau )d\tau +1.   
\end{align}
Performing the integration, we finally obtain
\begin{align}
S\left(  \Omega\right)   &  =\frac{\left\vert \gamma\right\vert ^{2}\Gamma_{sL}}%
{2v_{gs}\left[\Gamma_{sL}^2/4 + \Omega^2\right]}\left\{\left[  \left(  n_{th}+\frac{1}{2}\right)  \cosh\left[  2r\right]
-\frac{1}{2}\right]\right.  \nonumber\\
&  -\left. \left(  n_{th}+\frac{1}%
{2}\right)  \sinh\left(  2r\right)\cos(\phi-2\beta ) \right\} + 1.
\label{EqSGemeral}
\end{align}

Equation (\ref{EqSGemeral}) is the general result for the frequency-dependent squeezing in the channel waveguide. Perhaps of most interest is the squeezing at zero frequency, which is given simply by
\begin{align}
S\left(  0\right)  &=1+\frac{2\left\vert \gamma\right\vert ^{2}  }{v_{g}\Gamma_{sL}} 
\left[  \left(  n_{th}+\frac{1}%
{2}\right)  \cosh\left[  2r\right]\right.\nonumber \\
&\left. -\frac{1}{2} 
-\left(  n_{th}+\frac{1}%
{2}\right)  \sinh\left(  2r\right)  \cos\left(  2\beta-\phi\right)  \right].  
\end{align}

Taking the local oscillator phase to be $\beta=\phi (t_m)/2$ and using Eq. (\ref{Eqgamsq}), we obtain the following expression for the minimum in the squeezing in the waveguide:
\begin{equation}
S_{min}\left(  0\right)  =1+\left[1-\frac{Q_{sL}}{Q_{sI}}\right]\left[  \left(
\Delta X^{\min}\right)  ^{2}-1\right]  ,
\label{EqS0}
\end{equation}
where
\begin{equation}
\left(  \Delta X^{\min}\right)  ^{2}\equiv\left(  2n_{th}+1\right)  e^{-2r}%
\end{equation}
is the minimum quadrature variance inside the ring.

Note that for \textit{critical coupling} ($Q_{sI}=2Q_{sL}$), we obtain the usual result
\begin{equation}
S_{min}^{cc}\left(  0\right)   =\frac{1}{2}+\frac{1}{2}\left(  \Delta X^{\min}\right)  ^{2},    
\end{equation}
which shows that for critical coupling, the maximum squeezing is only 3 dB. In the limit of \textit{strong over-coupling} ($Q_{sI} \gg 2Q_{sL}$), we obtain%
\begin{equation}
S_{min}^{oc}\left(0\right)  =\left(  \Delta X^{\min}\right)  ^{2}.    
\end{equation}
Since the $Q_{sI}$ is limited by the intrinsic losses in the rings, in practice one has a trade off between the squeezing that can be obtained in the channel and the generation efficiency. Indeed, a lower loaded quality factor implies a weaker field enhancement inside the resonator and the need for a stronger pump. For a given platform, depending on the ring intrinsic loss and the available pump power, one needs to optimize the coupling between the ring and the waveguide to strike the balance between strong squeezing in the ring and strong coupling to the channel. This is what we do in the next section.

\section{Results} \label{Results}

In this section, we examine the effects of the different system parameters on the squeezing spectrum in the channel waveguide. The key system parameters are the pump duration ($\tau_{p}$) and strength ($g_0$), and the intrinsic and loaded $Q$ of the ring resonator at the signal frequency ($Q_{sI}$ and $Q_{sL}$, respectively) and pump frequency ($Q_{pI}$ and $Q_{sp}$, respectively).\\

Now, it is clear that the squeezing is always improved by increasing the intrinsic $Q$ of the ring at the signal frequency.  Moreover, we find that the best results can also be achieved for large intrinsic $Q$ for the pump. Thus, in all that follows, unless explicitly stated, we take $Q_{sI}=2\times 10^{6}$ and $Q_{pI}=8\times 10^{5}$, which are chosen to be close to the values found by Luo \textit{et al.} for a thin-film Lithium Niobate disk with $R=45 \mu m$. Note that these are conservative values relative to the best values in the literature of almost $3\times 10^6$ at the pump wavelength for an $R=100\mu m$ Lithium Niobate ring and almost $10^7$ at the signal for a $R=20\mu m$ Lithium Niobate disk \cite{Xie2021}. For simplicity, in what follows, we take $R=50 \mu m$, $\lambda_s=1550 nm$, and assume that the SPDC interaction is phase matched such that $n_{eff}(\omega_p)=n_{eff}(\omega_s)$, with $=n_{eff}=2.2$. This gives $T_{Rs}=T_{Rp}\equiv T_R \approx 2.3 ps$. Taking $\chi^{(2}_{eff}=54$ pm/V and $A_{eff}=0.71\mu m^2$, we find that with typical values of $g_0=1$ and $\widetilde{\tau}_p=2$, the peak electric field is less than 100 kV/cm. Therefore, the pump field values are small relative to damage thresholds or the values that would result in significant self-phase and cross-phase modulation. Moreover, the peak number of photons in the ring is typically about $10^6$; as we shall see, under desirable operating conditions, the number of signal photons will be much less than this, which means that pump depletion is not an issue either. With these values fixed, there are only four parameters with which to optimize the squeezing in the channel: 
\begin{align}
    f_s&\equiv Q_{sL}/Q_{sI},\\
    f_p&\equiv Q_{pL}/Q_{pI},
\end{align}
$\widetilde{\tau}_{p}$, and  $g_0$.\\

\begin{figure}[hbt]
\includegraphics[width=\columnwidth]{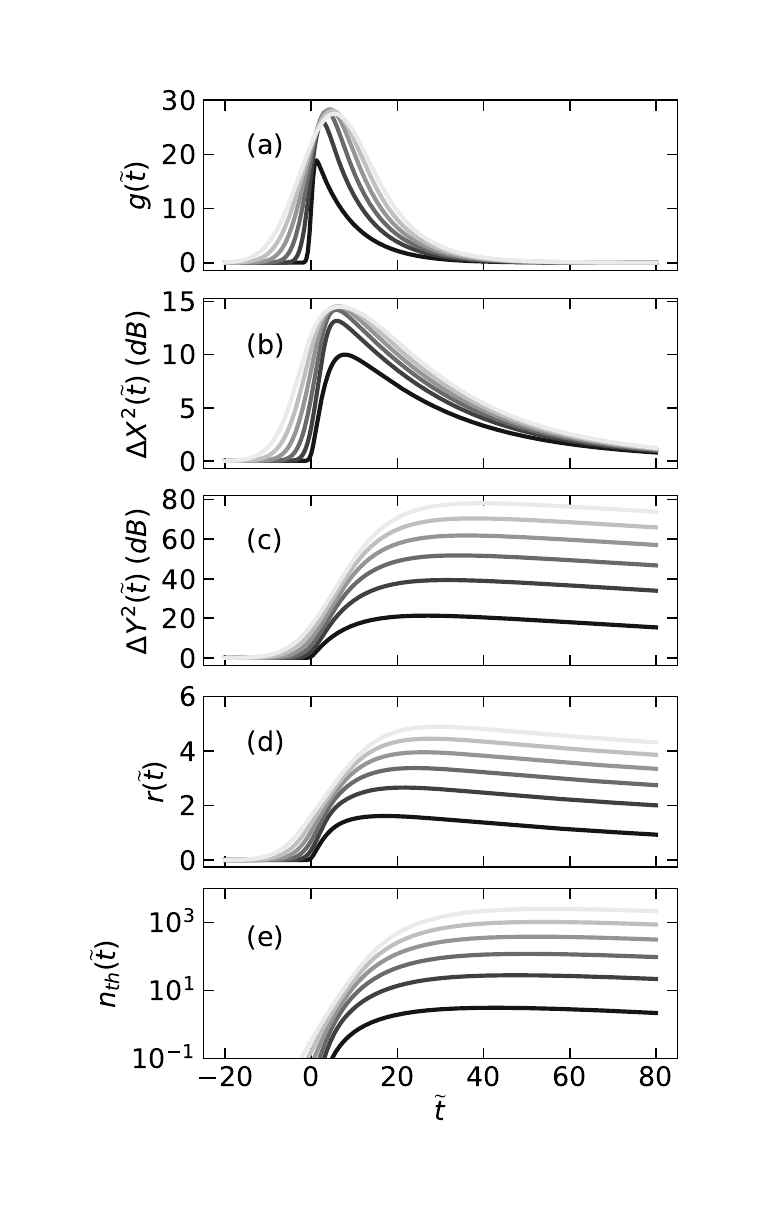}
\caption{The evolution of the signal light inside the ring resonator as a function of time. The six different curves are for values of the pulse duration, $\widetilde{\tau}_{p}=\{1,3,5,7,9,11\}$, with the darkest curve being $\widetilde{\tau}_{p}=1$.  The five plots are: (a) $g(\widetilde{t})$, (b) $\Delta X^2\left(\widetilde{t}\right)$, (c) $\Delta Y^2\left(\widetilde{t}\right)$, (c) $r\left(\widetilde{t}\right)$, and $n_{th}\left(\widetilde{t}\right)$. All plots are for $f_p=0.03$, $f_s=0.045$, and $g_0=1.0$. }\label{fig:gfuncplot}
\end{figure}
We start by presenting in Fig. \ref{fig:gfuncplot} plots of $g(\widetilde{t})$, $\Delta X^2\left(\widetilde{t}\right)$, $\Delta Y^2\left(\widetilde{t}\right)$ $r\left(\widetilde{t}\right)$, and $n_{th}\left(\widetilde{t}\right)$ as a function of time for $f_p=0.03$, $f_s=0.045$, and $g_0=1.0$ for six different values of $\widetilde{\tau}_{p}$. With these parameters, the number of pump photons in the channel waveguide is initially approximately $N_c\approx 7\times 10^5$ and the peak number of photons in the ring is about a quarter of that (depending on $\widetilde{\tau}_p$. Note that as the pulse duration increases, the maximum in $r\left(\widetilde{t}\right)$ increases monotonically. However, the maximum squeezing occurs for $\widetilde{\tau}_{p}\approx 7$. This is because for longer pulse durations, the number of thermal photons increases dramatically, leading to an eventual decrease in the peak squeezing.  Note, however, that the maximum squeezing value is relatively insensitive to the pump duration for $\widetilde{\tau}_{p}> 7$ but that the peak in the antisqueezing, $\Delta Y^2$ increases monotonically with $\widetilde{\tau}_{p}$ and is roughly four orders of magnitude larger for $\widetilde{\tau}_{p}=11$ than it is for $\widetilde{\tau}_{p}=1$. Therefore, as we shall see, it is preferable to use a shorter but more intense (larger $g_0$) pulse, if possible, to avoid large antisqueezing. Note also that the peak in the antisqueezing occurs at a much later time than the time at which the squeezing peaks. This has implications on the antisqueezing that one would measure in the channel waveguide, as we shall discuss shortly.\\

\begin{figure}[hbt]
\includegraphics[width=\columnwidth]{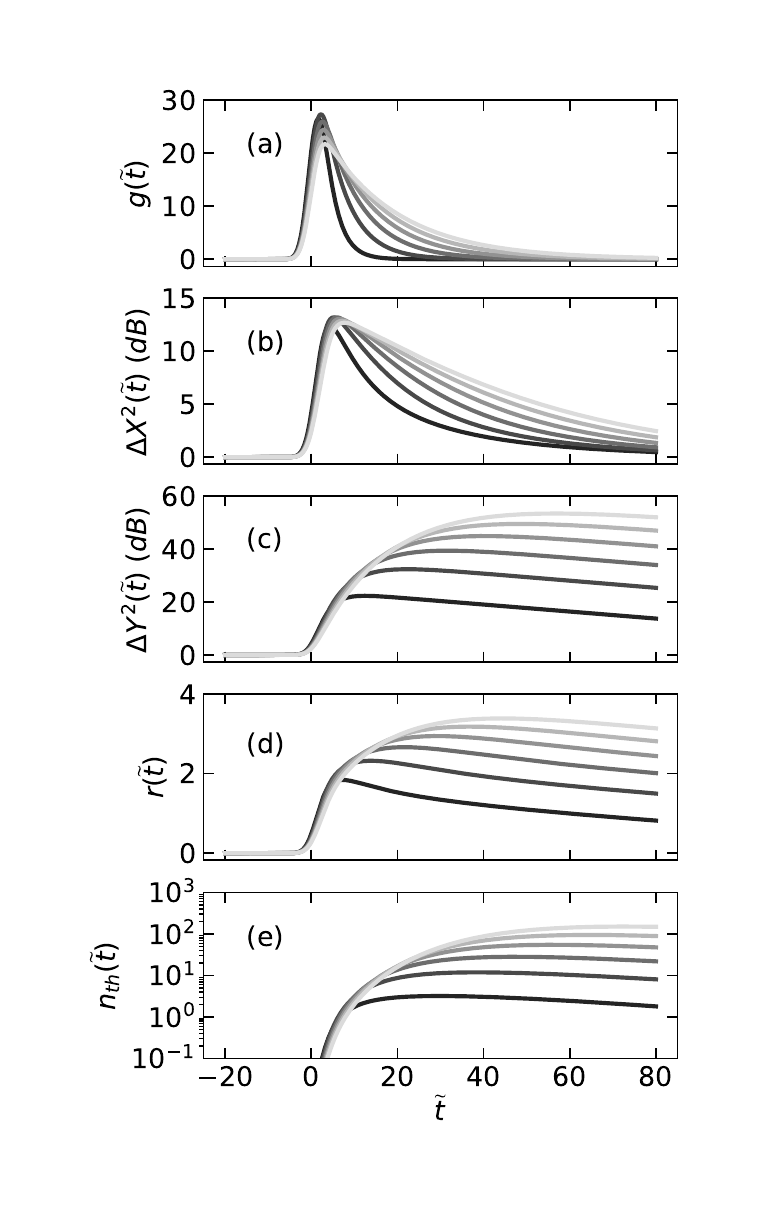}
\caption{The evolution of the signal light inside the ring resonator as a function of time. The six different curves are for values of the $Q$-ratio for the pump, $f_p=\{0.01,0.02,0.03,0.04,0.05,0.06\}$, with the darkest curve being $f_p=0.01$.  The five plots are: (a) $g(\widetilde{t})$, (b) $\Delta X^2\left(\widetilde{t}\right)$, (c) $\Delta Y^2\left(\widetilde{t}\right)$, (c) $r\left(\widetilde{t}\right)$, and $n_{th}\left(\widetilde{t}\right)$. All plots are for $\widetilde{\tau}_p=3$, $f_s=0.045$, and $g_0=1.0$. }\label{fig:gfuncplot2}
\end{figure}

The coupling of the pump field from the ring to the channel waveguide also has an important effect on the generated light.  This is because, when there is stronger coupling, the pump light will leave the ring more quickly, resulting in fewer thermal photons being generated.  Thus, it has a similar effect to changing the pump duration. To see this effect, in Fig. \ref{fig:gfuncplot2}, we plot $g(\widetilde{t})$, $\Delta X^2\left(\widetilde{t}\right)$, $\Delta Y^2\left(\widetilde{t}\right)$, $r\left(\widetilde{t}\right)$, and $n_{th}\left(\widetilde{t}\right)$ as a function of time for $\widetilde{\tau}_p=3$, $f_s=0.045$, and $g_0=1.0$, for 6 different values of $f_p$. As expected, increasing the loaded $Q$ for the pump increases the dwell time of the pump in the ring, which leads to a dramatic increase in the maximum squeezing parameter.  However, because it also leads to a dramatic increase in the thermal photon number, it has a relatively weak impact on the maximum squeezing for $f_p>0.03$. However, it has a very large effect on the maximum antisqueezing over the entire range. Thus, again there is a trade-off between maximizing the squeezing and keeping the antisqueezing at acceptable levels. This will be our main objective in what follows. 

We now turn to the problem of optimizing the squeezing of the light in the \textit{channel} waveguide. If one is willing to increase the amplitude of the input pulse amplitude sufficiently, then it is possible to achieve values for $S_{min}\left(0\right)$ that are greater than 15 dB.  However, if the pump level is too high, it can lead to other, parasitic, nonlinear processes that we have not taken into account here. In addition, when the squeezing is very high, it generally leads to very high antisqueezing, which can lead to problems if there is any significant error in the phase of the local oscillator used to measure the squeezing \cite{Vendromin2020}.

As can be seen from Figs. \ref{fig:gfuncplot} and \ref{fig:gfuncplot2}, the antisqueezing in the ring continues to grow for some time after the squeezing has maximized.  Therefore, using the level of antisqueezing in the ring at the time of maximum squeezing to calculate the antisqueezing in the channel waveguide will clearly underestimate the noise that would actually be measured.  Therefore, to obtain a better estimate of the maximum antisqueezing in the Channel, we first calculate the antisqueezing in the ring at the time $t_A$ at which the antisqueezing is a maximum.  We then assume that the antisqueezing takes that level from time $t_m$ to time $t_A$, after which it decays as the squeezing does.  Using this, the conservative estimate for the maximum antisqueezing in the ring for $\Omega =0$ is 
\begin{align}
S_{max}\left(  0\right)  =&1+\left[1-\frac{Q_{sL}}{Q_{sI}}\right]\left[1+\frac{\Gamma_{sL}(t_A - t_m)}{2}\right]\nonumber\\
&\times\left[  \left(
\Delta Y^{\max}\right)  ^{2}-1\right]  ,
\label{EqSMax}
\end{align}
where
\begin{equation}
\left(  \Delta Y^{\max}\right)  ^{2}\equiv\left[2n_{th}(t_A)+1\right]e^{2r(t_A)}%
\end{equation}
is the maximum in the antisqueezing variance inside the ring at time $t_A$. Note that because we have assumed that the antisqueezing is constant at its maximum value between the times $t_m$ and $t_A$, this is an overestimate of the maximum noise and 
\textit{we are guaranteed that the actual antisqueezing will always be less than this}.

\begin{figure}[hbt]
\includegraphics[width=\columnwidth]{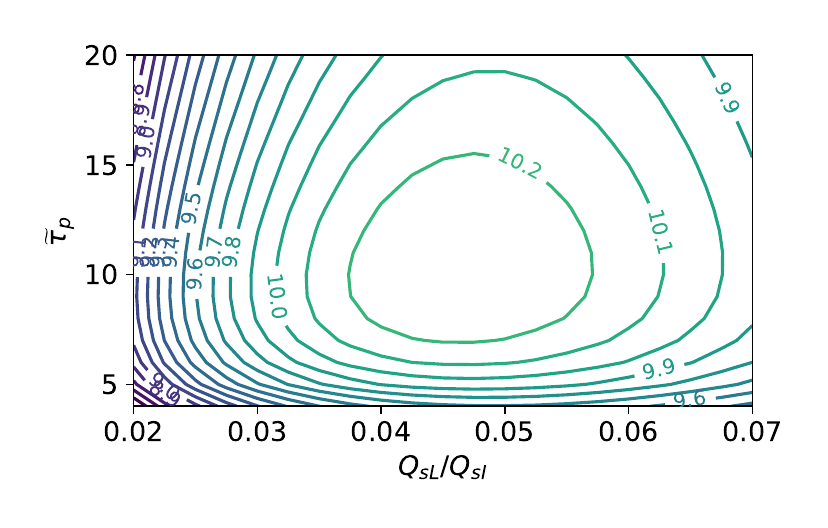}
\caption{A contour plot of the minimum value, $S_{min}(0)$ of the noise in the channel at zero frequency as a function of $f_s\equiv Q_{sL}/Q_{sI}$ and $\widetilde{\tau}_p$ for $f_p=0.03$, and $g_0=0.7$. The noise level is indicated in dB on the plot contours}\label{fig:Sminplot1}
\end{figure}

\begin{figure}[hbt]
\includegraphics[width=\columnwidth]{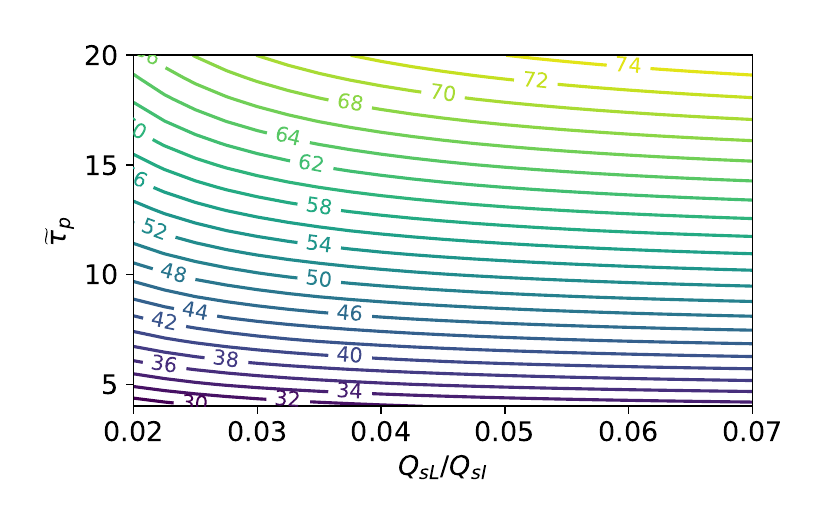}
\caption{A contour plot of the maximum value, $S_{max}(0)$ of the noise in the channel at zero frequency as a function of $f_s\equiv Q_{sL}/Q_{sI}$ and $\widetilde{\tau}_p$ for $f_p=0.03$, and $g_0=0.7$. The noise level is indicated in dB on the plot contours}\label{fig:SMaxplot1}
\end{figure}

In Fig. \ref{fig:Sminplot1}, we plot $S_{min}(0)$ as a function of $f_s$ and $\widetilde{\tau}_p$ for $g_0=0.7$, $f_p=0.03$.  As can be seen, the best squeezing occurs for $f_s\approx 0.48$ and $\widetilde{\tau}_p \approx 10$. For these values, $S_{min}(0)\approx 10.2$ dB.  However, the peak around the maximum is rather broad and there is a considerable range of values for $\widetilde{\tau}_p$ and $f_s$ over which the squeezing is greater than 10 dB.  To determine what might be the best choice of parameters, in Fig. \ref{fig:SMaxplot1}, we plot the maximum quadrature noise in the channel ($S_{max}(0)$ from Eq. (\ref{EqSMax})) as a function of the same two parameters.  As can be seen, the maximum noise is relatively insensitive to $f_s$, but is strongly dependent on $\widetilde{\tau}_p$.  If we assume that the desired level of squeezing is $S_{min}(0)= 10$ dB, then for this choice of $g_0$ and $f_p$, we see that the optimum set of parameters is $f_s\approx 0.05$ and $\widetilde{\tau}_p \approx 6$ for which $S_{max}(0)\approx 36$ dB.

\begin{figure}[hbt]
\includegraphics[width=\columnwidth]{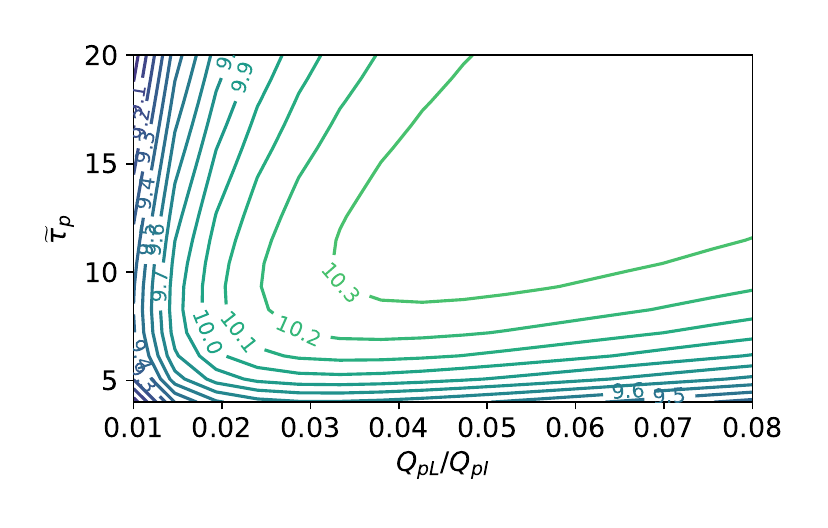}
\caption{A contour plot of the minimum value, $S_{min}(0)$ of the noise in the channel at zero frequency as a function of $f_p$ and $\widetilde{\tau}_p$ for $f_s=0.05$, and $g_0=0.7$. The noise level is indicated in dB on the plot contours}\label{fig:Sminplot2}
\end{figure}

\begin{figure}[hbt]
\includegraphics[width=\columnwidth]{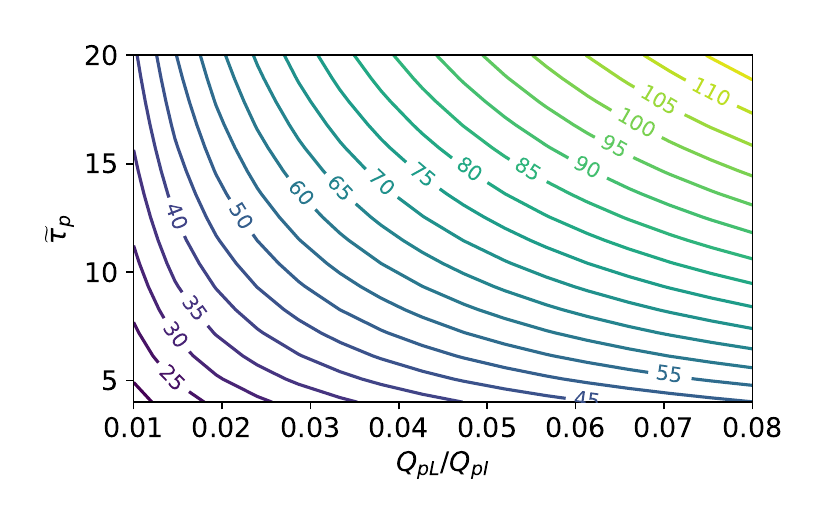}
\caption{A contour plot of the maximum value, $S_{max}(0)$ of the noise in the channel at zero frequency as a function of $f_p$ and $\widetilde{\tau}_p$ for $f_s=0.05$, and $g_0=0.7$. The noise level is indicated in dB on the plot contours}\label{fig:SMaxplot2}
\end{figure}

Given the sensitivity of the antisqueezing to the pump duration, one would expect a similar sensitivity to $f_p$, as both of these parameters control the duration of the pump pulse in the ring.  In Figs. \ref{fig:Sminplot2} and \ref{fig:SMaxplot2}, we plot the squeezing and antisqueezing, respectively, as a function of $\widetilde{\tau}_p$ and $f_p$, for $g_0$=0.7 and $f_s=0.05$. As can be seen, both the squeezing and antisqueezing decrease as you decrease either of the two plot parameters.  However, the decrease in the antisqueezing is much more dramatic.  

\begin{figure}[hbt]
\includegraphics[width=\columnwidth]{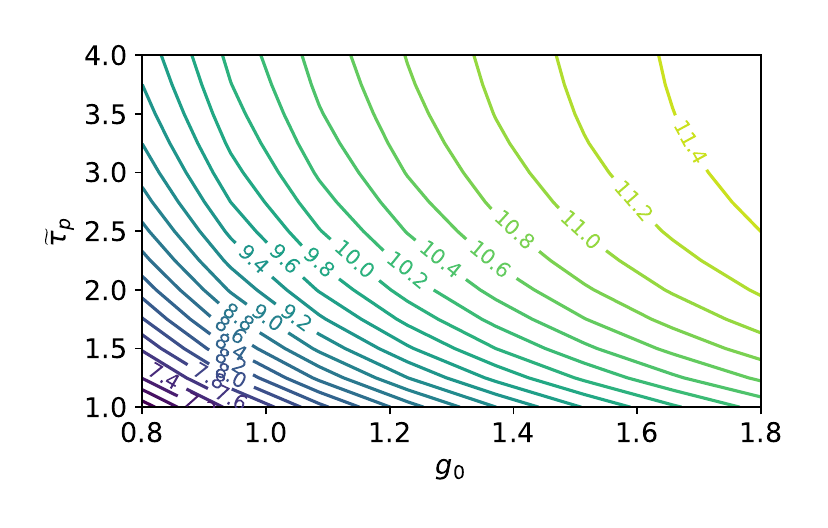}
\caption{A contour plot of the minimum value, $S_{min}(0)$ of the noise in the channel at zero frequency as a function of $g_0$ and $\widetilde{\tau}_p$ for $f_s=0.05$, and $f_p=0.01$. The noise level is indicated in dB on the plot contours}\label{fig:Sminplot3}
\end{figure}

\begin{figure}[hbt]
\includegraphics[width=\columnwidth]{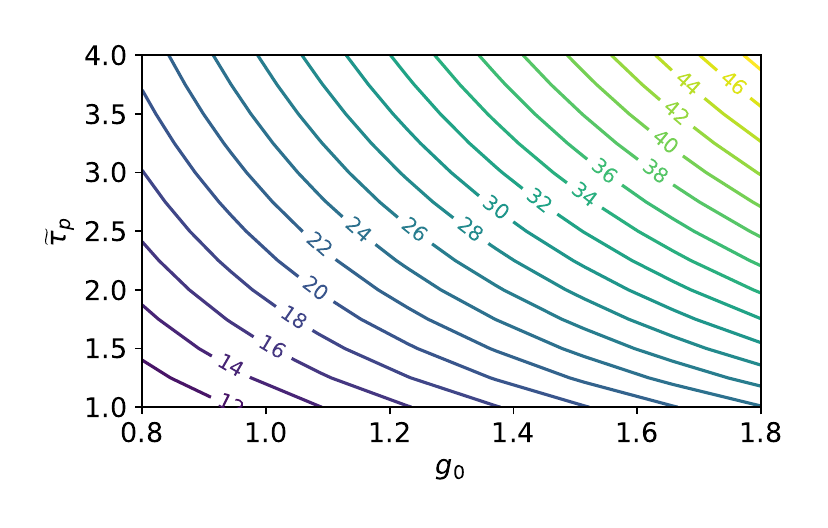}
\caption{A contour plot of the maximum value, $S_{max}(0)$ of the noise in the channel at zero frequency as a function of $g_0$ and $\widetilde{\tau}_p$ for $f_s=0.05$, and $f_p=0.01$. The noise level is indicated in dB on the plot contours}\label{fig:SMaxplot3}
\end{figure}

The results presented in Figs. \ref{fig:Sminplot1} to \ref{fig:SMaxplot2} indicate that, if possible, one should minimize both the pulse duration and the loaded $Q$ for the pump to minimize the antisqueezing. We can still achieve our target squeezing of 10 dB as long as we increase $g_0$ sufficiently to compensate for the high loss in the ring. In Figs. \ref{fig:Sminplot3} and \ref{fig:SMaxplot3}, we plot the squeezing and antisqueezing, respectively, as a function of $\widetilde{\tau}_p$ and $g_0$, for $f_p$=0.01 and $f_s=0.05$. For this subset of parameters, the optimal point to minimize the antisqueezing, while achieving a squeezing of 10 dB, is $\widetilde{\tau}_p$=1.0 and $g_0$=1.68, which gives $S_{max}(0)\approx 23$ dB.

We now search for the globally optimal system that achieves 10 dB of squeezing while minimizing the antisqueezing. It is clear that the optimal results will be achieved for as low a loaded $Q$ for the pump as possible.  However, achieving high squeezing for low-$Q_pL$ structures requires more energy in the incident pump.  Moreover, our approximate analytic form for the pump in the ring becomes less accurate if $f_p$ is too small.  For example, for $f_p=0.01$ and $\widetilde{\tau}_p=1$, although we obtain an accurate pulse duration in the ring, the peak value is only 80\% of the analytic one. Furthermore, the pump pulse duration cannot go below $\widetilde{\tau}_p=1$ or there will be coupling into other ring modes \cite{Vendromin2020}. Therefore, in our search for the optimal configuration, we restrict the parameters as follows: $\widetilde{\tau}_p\geq 1$ and $f_p \geq 0.01$. With these constraints, we find that the optimal values for the parameters are $g_0=1.7$, $f_p$=0.01, $\widetilde{\tau}_p=1.0$, and $f_s=0.03$, for which we obtain a maximum antisqueezing in the channel of only 21.9 dB.  We note also that for this optimum configuration, the total number of signal photons generated (including those lost) is only 62, meaning that the undepleted pump approximation is well justified. 

\section{Conclusion} \label{conclusion}

In this work, we developed and applied a semianalytic formalism to model and optimize pulsed squeezed state generation in a channel waveguide side-coupled to a lossy ring resonator system. Our approach provides a robust method to analyze the evolution of quadrature squeezing under realistic conditions while accounting for the key physical parameters systems, such as size and loaded and intrinsic quality factors.

We began by deriving an approximate yet accurate expression for the pump field within the ring resonator in the presence of a temporally Gaussian pump pulse. This allowed us to calculate the evolution of the signal density operator, identifying the critical time at which the quadrature squeezing inside the ring is maximized. Leveraging a (conservative) assumption that the pump contribution to squeezing can be neglected beyond this point, we determined the free evolution of the system. This allowed us to find semianalytic expressions for the squeezing spectrum measurable in the output channel waveguide.

We found the optimal parameters for our system that correspond to a compromise between strong squeezing in one quadrature and minimal antisqueezing in the orthogonal quadrature. Specifically, we showed that achieving a squeezing level above 10 dB in the channel is feasible, provided that (1) the loaded quality factor of the signal mode is maintained at approximately 3\% of the intrinsic quality factor, (2) the pump pulse duration is kept as short as possible while maintaining single-mode coupling, and (3) the loaded quality factor of the pump mode is minimized to reduce thermal signal photon contributions to the noise.

By systematically exploring our parameter space, we established the critical trade-offs between squeezing and antisqueezing. For instance, while longer pump pulses or higher-loaded pump Q-factors enhance squeezing, they also lead to significant antisqueezing due to increased thermal noise. We showed that a short, intense pump pulse combined with a low-loaded Q for the pump mode offers an optimal balance. Under these conditions, we showed that a 10 dB squeezing level can be achieved with antisqueezing levels remaining below 22 dB.

Our results confirm the importance of critically optimizing the coupling between the ring resonator and the waveguide. Over-coupling reduces squeezing within the ring, while under-coupling limits the amount of squeezed light extracted into the channel. The results we achieved clarify the interplay between system losses, nonlinear coupling, and temporal pump characteristics.

These findings are significant not only for their direct implications in integrated quantum photonics but also for their broader relevance to the development of scalable and efficient sources of squeezed states for quantum technologies. Future work could include extending our formalism to account for pump depletion effects and parasitic nonlinear processes, as well as experimental validation of our theoretical predictions.

\textit{Acknowledgements.}
M.M.D. acknowledges the support of the Canada Foundation for Innovation and the Natural Sciences and Engineering Research Council of Canada (NSERC). M.L acknowledges the support of PNRR MUR project PE0000023-NQSTI.



\bibliographystyle{apsrev4-2}
\bibliography{main}

\end{document}